\documentstyle[11pt,aaspp4]{article}

\slugcomment{Submitted to ApJ}

\begin{document}

\title{On the Neutrino Flux from Gamma-Ray Bursts}

\author{D. Guetta\altaffilmark{1}, M. Spada\altaffilmark{1},
E. Waxman\altaffilmark{2}}
\altaffiltext{1}{Osservatorio di Arcetri Largo E. Fermi 5, 50125 Firenze,
Italy}
\altaffiltext{2}{Department of Condensed Matter Physics, Weizmann
Institute, Rehovot 76100, Israel}

\begin{abstract}

Observations imply that gamma-ray bursts (GRBs) are produced by the
dissipation of the kinetic energy of a highly relativistic fireball.
Photo-meson interactions of protons with $\gamma$-rays
within the fireball dissipation region are expected to convert 
a significant fraction of fireball energy to $>10^{14}$~eV neutrinos.
We present an analysis of the internal shock model of GRBs, where
production of synchrotron photons and photo-meson neutrinos
are self-consistently calculated, and
show that the fraction of fireball energy converted 
to high energy neutrinos is not sensitive to uncertainties
in fireball model parameters, such as the expansion Lorentz factor
and characteristic variability time. This is due in part to the constraints 
imposed on fireball parameters by observed 
GRB characteristics, and in part to the fact that for 
parameter values for which the photo-meson optical depth is high 
(implying high proton energy loss to pion production) neutrino production
is suppressed by pion and muon synchrotron losses.
The neutrino flux is therefore
expected to be correlated mainly with the observed $\gamma$-ray flux.
The time averaged neutrino intensity predicted by the model, 
$\sim10^{-8.5}{\rm GeV/cm}^2{\rm s\,sr}$, is consistent with the flux 
predicted by the assumption that GRBs are the sources of $>10^{19}$~eV
cosmic-rays.

\end{abstract}


\section{INTRODUCTION}

The characteristics of $\gamma$-ray bursts (GRBs),
bursts of 0.1 MeV--1 MeV photons lasting for a few seconds
(\cite{Fishman}), suggest that the observed radiation is produced by 
the dissipation of the kinetic energy
of a relativistically expanding wind, a 
``fireball,'' at cosmological distance
(see, e.g., \cite{fireballs} for review). 
The recent detection of delayed low energy (X-ray to radio) emission 
(afterglow) from GRB sources (see \cite{AG_ex_review} for review),
confirmed both the cosmological origin of the bursts and 
standard model predictions of afterglows,
that result from the collision of an expanding fireball with
its surrounding medium (see \cite{AG_th_review} for review). 

Within the fireball model framework, observed $\gamma$-rays are produced
by synchrotron emission of electrons accelerated to high energy by internal
shocks within the expanding wind. In the region where electrons are 
accelerated, protons are also expected to be shock accelerated: 
Plasma parameters in the dissipation region allow proton acceleration 
to $>10^{20}$~eV (\cite{W95a,Vietri95};
see \cite{My_revs} for a recent review). A natural
consequence of proton acceleration to high energy is the production of
a burst of $\gtrsim10^{14}$~eV neutrinos (\cite{WnB97},2000),
produced by the decay of charged pions created in interactions between
fireball photons and high energy protons.  Lower energy neutrinos
may be produced by inelastic nuclear collisions (\cite{BnM00,MnR00}).

The characteristic energy of neutrinos produced by 
photo-meson interactions is determined by the relation 
between the observed photon energy, $E_\gamma$,
and the accelerated proton's energy, $E_p$,
at the photo-meson threshold  of the $\Delta$-resonance.
In the observer frame,
\begin{equation}
E_\gamma \, E_{p} = 0.2 \, {\rm GeV^2} \, \Gamma^2,
\label{eq:keyrelation}
\end{equation}
where phenomenologically the Lorentz factors of the
expanding fireball are $\Gamma > 10^{2}$.
For typical observed $\gamma$-ray energy of 1~MeV,
proton energies $E_p\approx 2\times10^7$~GeV are 
required to produce neutrinos
from pion decay, leading to $\approx 10^{15}$~eV neutrinos.
Both $\sim1$~MeV photons and $\sim10^{15}$~eV neutrinos are produced in
this model during the stage of internal shocks within the expanding wind.
Much higher energy, $\gtrsim 10^{18}$~eV, neutrinos may be produced at a later
stage, at the onset of fireball interaction with its surrounding medium. 
Optical--UV photons are radiated  by electrons accelerated in shocks
propagating backward into the fireball plasma.
Protons are accelerated to high energy
in these ``reverse'' shocks.
The combination of low energy photons and high energy protons produces
ultra-high energy neutrinos via photo-meson interactions, as indicated by
Eq. (\ref{eq:keyrelation}).

The predicted flux of $\gtrsim10^{14}$~eV neutrinos 
produced in internal shocks 
is determined by the fraction $f_\pi$ of fireball 
proton energy lost to pion production. This fraction is determined by the
number density of photons at the internal dissipation region, and is given by
(\cite{WnB97},1999)
\begin{equation}
f_\pi(E_p)\approx0.2\min(1,E_p/E_p^b){L_{\gamma,52}\over
\Gamma_{2.5}^4 \Delta t_{-2}E_{\gamma,\rm MeV}^b}.
\label{eq:fpi}
\end{equation}
The $\gamma$-ray luminosity $L_\gamma$, 
the photon spectral break energy $E_\gamma^b$ (where 
luminosity per logarithmic photon energy interval peaks),
the observed variability time $\Delta t$
and the wind Lorenz factor $\Gamma$ 
are normalized in Eq. (\ref{eq:fpi}) to their
typical values inferred from observations:
$L_{\gamma}=10^{52}{\rm erg/s}$, $E_{\gamma}^b=1E_{\gamma,\rm MeV}^b$~MeV, 
$\Delta t=10^{-2}$s,
and $\Gamma=10^{2.5}$. The proton break energy, $E_p^b$, is the threshold
proton energy for interaction with photons of observed energy
$E_{\gamma}^b$,
$E_p^b=(2/E_{\gamma,\rm MeV}^b)\Gamma_{2.5}^2\times10^7$~GeV.

The rate at which energy is produced as $\gamma$-rays by GRBs is 
similar to the rate of production of high energy, $>10^{19}$~eV, protons
implied by the observed ultra-high energy cosmic-ray flux 
(\cite{W95a},b,2000), 
\begin{equation}
E_p^2{d\dot n_p\over dE_p}\approx
10^{44}{\rm erg\ Mpc}^{-3}{\rm yr}^{-1}.
\label{eq:ECR}
\end{equation}
Assuming that GRBs are the sources of observed 
ultra-high energy cosmic rays, the intensity of high energy
muon neutrinos and anti-neutrinos
implied by Eq. (\ref{eq:fpi}) is then (\cite{WnB97},1999)
\begin{equation}
E_\nu^2\Phi_{\nu}\approx
3\times10^{-9}\left[{f_\pi(E_p>E_p^b)\over0.2}\right]
\min(1,{E_\nu/E^b_\nu}) {\rm GeV/cm^2\,sr\,s}.
\label{eq:JGRB}
\end{equation}
Since neutrinos carry $\approx5\%$ of the proton energy, the neutrino break
energy is given by
\begin{equation}
E_\nu^b\approx10^{15}(1+z)^{-2}{\Gamma_{2.5}^2\over
E_{\gamma,\rm MeV}^b}{\rm eV}.
\label{eq:Enub}
\end{equation}
We have explicitly introduced here the 
dependence on source redshift, $z$, which 
is due to the fact 
that energies observed on Earth are smaller than those measured at the
source redshift by a factor $(1+z)$.

The value of $f_\pi$, Eq. (\ref{eq:fpi}), is strongly dependent on 
$\Gamma$. It has recently been pointed out by Halzen
and Hooper (1999) that if the Lorentz factor $\Gamma$ varies significantly
between bursts, with burst to burst variations $\Delta\Gamma/\Gamma\sim1$, 
then the resulting neutrino
flux will be dominated by a few neutrino bright bursts, and may 
significantly exceed 
the flux given by Eq. (\ref{eq:JGRB}), derived 
for typical burst parameters.
This may strongly enhance the detectability of GRB neutrinos
by planned neutrino telescopes (\cite{Alvarez00}).

The main goal of the present work is to determine the allowed range of
variation in the 
fraction of fireball energy converted to high energy neutrinos, under the 
assumption that GRBs are produced
by internal dissipation shocks in an ultra-relativistic wind. We 
self-consistently calculate
the dependence on wind parameters of both $\gamma$-ray emission by electrons 
and neutrino production by photo-meson interaction of protons,
taking into account synchrotron losses of high energy pions and muons
and pair-production interaction of high energy photons. 
We show that large burst-to-burst variations in the fraction of
fireball energy converted to neutrinos are not expected, due to two reasons.
First, the observational constraints imposed by $\gamma$-ray observations,  
in particular the requirement $E_{\gamma}^b\ge0.1$~MeV, 
imply that wind model
parameters ($\Gamma$, $L$, $\Delta t$) are correlated, and that
$\Gamma$ is restricted to values
in a range much narrower than $\Delta\Gamma/\Gamma\sim1$. 
Second, for wind parameters that imply, through Eq. (\ref{eq:fpi}),
$f_\pi$ values significantly exceeding 20\%, 
only a small fraction of the pions' energy is converted
to neutrinos due to pion and muon synchrotron losses. 

The model we analyze is described in \S~\ref{sec:model}. Numerical results are
presented and discussed in \S~\ref{sec:results}. 
Conclusions and implications
are summarized in \S~\ref{sec:discussion}.

\section{OUTLINE OF THE MODEL}
\label{sec:model}

We consider a compact source, of linear scale
$R_0\sim10^6$~cm, which produces a wind characterized by an average luminosity 
$L_w\sim10^{52}{\rm erg\,s}^{-1}$--$10^{53}{\rm erg\,s}^{-1}$
and mass loss rate $\dot M=L_w/\eta c^2$
($R_0\sim10^6$~cm corresponds to three times the Schwarzschild radius of
a non-rotating, solar mass black hole). At small radius, 
the wind bulk Lorentz factor, $\Gamma$, 
grows linearly with radius, until most of the wind energy is converted
to kinetic energy and $\Gamma$ saturates at $\Gamma\sim\eta\sim300$.
$\eta\gtrsim300$ is required to reduce the wind pair-production
optical depth for observed
high energy, $>100$~MeV, photons to less than unity.
If $\eta>\eta_*\approx(\sigma_T L_w/4\pi m_p c^3 R_0)^{1/4}=
3\times10^3(L_{w,53}/R_{0,6})^{1/4}$, where 
$L_w=10^{53}L_{w,53}{\rm erg\,s}^{-1}$ and $R_0=10^6R_{0,6}$~cm,
the wind becomes optically thin at $\Gamma\approx\eta_*<\eta$, and hence
acceleration saturates at $\Gamma\approx\eta_*$ and the remaining wind
internal energy escapes as thermal radiation at $\sim1$~MeV temperature.
Variability of the source on time scale $\Delta t$, resulting
in fluctuations in the wind bulk Lorentz factor $\Gamma$ on similar
time scale, leads to internal shocks
in the expanding fireball at a radius
$R_i\approx2\Gamma^2c \Delta t=6\times10^{13}\Gamma^2_{2.5} \Delta t_{-2}
{\rm\ cm}$.
If the Lorentz factor variability within the wind is significant,
internal shocks reconvert a substantial 
part of the kinetic energy to internal energy. It is assumed that
this energy is then radiated as 
$\gamma$-rays by synchrotron (and inverse-Compton) emission of
shock-accelerated electrons.

The dynamics of our model is described in \S~\ref{sec:dynamics}, and
neutrino production calculations are described in \S~\ref{sec:nus}.

\subsection{Model Dynamics}
\label{sec:dynamics}

In this work, we use an approximate model of the
unsteady wind described in the preceding paragraph, following 
Spada, Panaitescu \& M\'esz\'aros 2000, and Guetta, Spada \& Waxman 2000
(GSW00). The wind evolution is
followed starting at radii larger than the saturation radius, i.e. 
after the shells have already reached their final 
Lorenz factor following the acceleration phase, and the GRB photon 
flux and spectrum resulting from a series 
of internal shocks that occur within the wind at larger radii
are calculated. 

The wind flow is approximated as a set of discrete shells. 
Each shell is characterized by four 
parameters: ejection time $t_j$, where the subscript $j$ denotes the
$j$-th shell, Lorentz factor $\Gamma_j$, mass $M_j$, and width $\Delta_j$. 
Since the wind duration, $t_w\sim10$~s, is much larger than the dynamical
time of the source, $t_d\sim R_0/c$, variability of the wind on a wide range
of time scales, $t_d<t_v<t_w$, is possible. For simplicity, we consider
a case where the wind variability is characterized by a single time scale 
$t_w>t_v>t_d$, in addition to the dynamical time scale of the source $t_d$
and to the wind duration $t_w$.
Thus, we consider shells of initial thickness 
$\Delta_j=c t_d=R_0$, ejected from the source
at an average rate $t_v^{-1}$. 

In GSW00 we have examined the dependence of observed $\gamma$-ray 
flux and spectrum on wind model parameters, taking into account
both synchrotron and inverse-Compton emission, 
and the effect of $e^\pm$ pair production. We have assumed there
that the Lorenz factor of a given shell is independent of 
those of preceding shells, and considered various Lorentz factor 
distributions.
We have shown that in order to obtain $\gamma$-ray flux and spectrum 
consistent with observations, large variance is required in wind Lorentz
factor
distribution. We therefore restrict the following discussion to 
the bimodal case, where Lorentz factors are drawn from a bimodal distribution,
$\Gamma_j=\Gamma_m$ or $\Gamma_j=\Gamma_{M}\approx\eta_*\gg\Gamma_m$, 
with equal probability.
The time intervals $t_{j+1}-t_j$ are drawn randomly from
a uniform distribution with an average value of $t_v$.
In GSW00 we have considered two qualitatively 
different scenarios for shell masses distribution,
shells of either equal mass or equal energy, and concluded that
observations favor shells of equal mass. We therefore restrict the following 
discussion to the equal shell mass case. We note, however, that our
conclusions are not sensitive to this assumption.

Once shell parameters are determined, we calculate the radii where  
collisions occur and determine the photon and neutrino emission 
from each collision. 
In each collision a forward and a reverse shock are formed,
propagating into forward and backward shells respectively.
The plasma parameters behind each shock are determined by the Taub
adiabatic, requiring continuous energy density and velocity across the contact 
discontinuity separating the two shells (Panaitescu \& M\'esz\'aros 1999).
We assume that a fraction $\epsilon_e$ ($\epsilon_B$) of the protons
shock thermal energy is converted to electrons (magnetic field).
We assume that both electrons and protons are accelerated by the shocks to a 
a power-law distribution, $dn_\alpha/d\gamma_\alpha\propto\gamma_\alpha^{-p}$ 
for particle
Lorentz factors $\gamma_{\alpha,\min}<\gamma_\alpha<\gamma_{\alpha,\max}$. 
The maximum Lorenz factor is determined by equating the acceleration
time, estimated as the Larmor radius divided
by $c$, to the minimum of the dynamical time and the synchrotron 
cooling time.

The calculation of emitted $\gamma$-ray spectrum and flux is carried
according to the method described in 
detail in GSW00, taking into account both 
synchrotron and inverse Compton emission, as well as the
effect of the optical thickness due to Thomson scattering on both 
electrons present initially in the flow and those created by pair
production. 
Here we focus on neutrino production by photo-meson interactions,
following Waxman \& Bahcall 1997. Our method of calculation is described below.

\subsection{Neutrino production}
\label{sec:nus}

The fraction $f_{\pi}(E_{p})$ of proton energy 
lost to pion production is estimated as
$f_{\pi}={\rm min}(1,\Delta t/t_{\pi})$ where $\Delta t$ is
the comoving shell expansion time and $t_{\pi}$ is the proton 
photo-pion energy loss time (Waxman \& Bahcall 1997),
\begin{eqnarray}
\label{eq:tpi}
t_{\pi}^{-1}(E_{p}) &=& -\frac{1}{E_{p}}
\frac{dE_{p}}{dt} \nonumber \\
&=& 
\frac{1}{2\gamma_{p}^{2}}c\int dE
\sigma_{\pi}(E) \xi(E) E
\int_{E/2\gamma_{p}}^{\infty}dx x^{-2} n(x).
\end{eqnarray}
Here $\sigma_{\pi}(E)$ is the cross section for the 
pion production for a photon with energy $E$ 
in the proton rest frame, $\xi(E)$ is the 
average fraction of energy lost to the pion, and
$n(E)$ is the comoving number density of photons per unit energy.
In our calculations, we have approximated the integral, Eq. (\ref{eq:tpi}),
by the contribution from the $\Delta$ resonance. This approximation is
valid for GRB photon spectra (\cite{My_revs}). 

Photo-meson production of low energy protons, well below
$E_p^b$, requires interaction with 
high energy photons, well above $E_\gamma^b$. Such photons
may be depleted by pair production. 
For each collision we find the photon energy 
$E_{\gamma}^{\pm}$, for which the pair production optical thickness,
$\tau_{\gamma\gamma}$, is unity. A large fraction of 
photons of energy exceeding $\max(E^{\pm}_{\gamma},m_e c^2)$ (measured
in the shell frame) 
will be converted
to pairs, and hence will not be available for photo-meson interaction,
leading to a suppression of the neutrino flux
at low energies. In order to take this effect into account,
we use  
$f_{\pi}={\rm min}[1,(\Delta t/t_{\pi})\min(1,\tau_{\gamma\gamma}^{-1})]$
for protons interacting with photons which are strongly suppressed by pair
production.

Neutrino production is suppressed at high energy,
where neutrinos are produced by the decay of muons and pions 
whose lifetime $\tau$ exceeds the characteristic 
time for energy loss due to synchrotron emission (Waxman \&
Bahcall 1997, 1999, Rachen \& M\'eszar\'os 1998).
We therefore define an effective $f_\pi$, $f_{\pi,\rm eff.}$, as 
$4f_\pi$ times the fraction of pions' energy converted to muon neutrinos.
In the absence of pion and muon energy loss, $\approx1/4$ of the pions'
energy is converted to muon neutrinos, 
since $\approx1/2$ the energy of charged pions is converted to 
$\nu_\mu+\bar\nu_\mu$. Thus, $f_{\pi,\rm eff.}$ is the fraction of proton
energy that, in the absence of synchrotron losses, leads to approximately 
the same muon neutrino flux as that obtained when synchrotron
losses are taken into account.

In each collision a fraction 
of the kinetic energy of the colliding shell  is converted to a flux 
of photons and neutrinos. The energy that is not lost to 
photons and neutrinos is converted back to kinetic energy by adiabatic
expansion of the shell.

\section{RESULTS AND DISCUSSION}
\label{sec:results}

In this section we determine the dependence of $f_{\pi,\rm eff.}$, 
the fraction of proton
energy that in the absence of synchrotron losses leads to approximately 
the same muon neutrino flux as that obtained when synchrotron
losses are taken into account,
on wind model parameters. We adopt electron and magnetic field energy 
fractions
close to equipartition, $\epsilon_e=0.45$ and $\epsilon_B\ge0.01$, and $p=2$. 
The equipartition fractions are required to satisfy
$\epsilon_e\ge0.1$ and $\epsilon_B\ge0.01$ in order to  
account for observed $\gamma$-ray emission (e.g. GSW00). 
Moreover, $\epsilon_e$ values in the range of 0.1 to 0.5 are typically
derived from GRB afterglow observations
(e.g. \cite{Freedman}). $p\simeq2$ is required to account
for observed $\gamma$-ray and afterglow spectra (\cite{Freedman}). 
Figure 1 presents contour plots of the photon break energy, $E_\gamma^b$,
as function 
of $\Gamma_m$ and $t_v$ for $L_w=10^{53}{\rm erg/s}$ and 
and $\epsilon_B=0.01$, $\epsilon_B=0.1$.
The qualitative behavior demonstrated by the contour plot
can be understood based on the following arguments. The wind energy
density at the smallest radii of the internal dissipation shocks 
is lower for higher values of
$\Gamma_m$, leading to a decrease of $E_\gamma^b$ with increasing
$\Gamma_m$. This effect sets the upper bound on $\Gamma_m$ values. Lowering
the value of $\Gamma_m$ leads to increase in $E_\gamma^b$ together with
increase in the wind optical depth at the smallest internal
shocks radii. Once the optical depth exceeds unity, emission is strongly 
suppressed leading to a rapid decrease with $\Gamma_m$ of both the 
fraction of wind energy escaping as radiation and $E_\gamma^b$ 
(see GSW00 for detailed discussion). 

Figures 2 presents the contour plot
for the effective $f_\pi$, $f_{\pi,\rm eff.}$, for the case shown in
Figure 1 with 
$\epsilon_B=0.01$. The values shown are averaged over all internal collisions, 
weighted by the energy converted to high energy protons in each collision. 
Values of $f_\pi$ are shown for several neutrino energy ranges, using
the approximation $E_\nu=0.05E_p$. The region in 
$\Gamma_m$--$t_v$ plane where $E_\gamma^b>0.1$~MeV
is bound in Figures 2 by the dashed line.
The dash-dotted line outlines the region in which 
the fraction of wind energy converted to radiation
exceeds 2\% (higher fraction is obtained for larger $\Gamma_m$ values).
For wind parameters
consistent with observed GRB characteristics,
the effective value of $f_\pi$ at high neutrino energy, $E_\nu>E_\nu^b$,
is restricted to values in the range of $\approx10\%$ to 
$\approx30\%$.
Figure 3 presents the fraction $f_{\nu}^{\rm thin}$
of the fireball neutrino flux produced by
internal shocks which are optically thin to radiation. While at high
energy most of the neutrino flux is produced in the optically thin 
regime, where collisions also contribute to the photon flux, a large
fraction, $\sim50\%$, of 
the neutrino flux at low energy is produced by collisions which
do not contribute to the observed gamma-ray flux. The optically thick
collisions provide a significant contribution only at low energy, since the
large energy density at these collisions lead to strong suppression of 
high energy neutrino production by muon and pion synchrotron losses.

In figure 
4 we present contour plots of $f_{\pi}$, neglecting the 
effects on neutrino production of synchrotron losses and pair-production.
$f_{\pi}$ approaches unity at low values of $\Gamma_m$ for low
proton (neutrino) energy, and at intermediate values of $\Gamma_m$
for high proton (neutrino) energy. 
The energy loss of pions and
muons due to synchrotron emission reduces the fraction of pion energy
converted to neutrinos, and hence suppresses the effective value of
$f_{\pi}$, to the range of values shown in Figure 2. The suppression
seen in Figure 4
of $f_\pi$ at low $\Gamma_m$ for high energy protons is due to the fact
that for low values of $\Gamma_m$ most collisions occur at small radii, 
where synchrotron losses of high energy protons prevent acceleration of 
protons to ultra-high energy (\cite{W95a}). Ultra-high energy protons are
produced in this case only at secondary collisions at large radii, where
the photon energy density is already low, resulting in smaller values of
$f_\pi$.

Figures 5 and 6 present $f_{\pi,\rm eff.}$ for $L_w=10^{53}{\rm erg/s}$,
$\epsilon_B=0.1$ and $L_w=10^{52}{\rm erg/s}$ 
$\epsilon_B=0.01$ respectively. Comparing to the results presented in
Figure 2, we find that the qualitative behavior as well as the predicted 
values of $f_{\pi,\rm eff.}$ are not sensitive to wind luminosity 
and magnetic field equipartition fraction (over the range allowed by GRB
observations).

The neutrino break energy inferred from Figures 2, 5 and 6,
$E_\nu^b\sim10^{15}$~eV, is somewhat higher than estimated by Waxman
\& Bahcall (1997), $E_\nu^b\sim10^{14}$~eV [see Eq. (\ref{eq:Enub})],
since the photon energy break in our wind models is typically 
$E_\gamma^b\sim0.1$~MeV, rather than $E_\gamma^b\sim0.5$~MeV. Note, that
although synchrotron losses suppress neutrino production at high energy
(compare Figures 2 and 4), the increase in $f_\pi$ at high energy (Figure
2) implies that the neutrino flux is flat, with roughly equal energy
per logarithmic neutrino energy interval above $E_\nu^b$, in 
contrast with the expectation of strong suppression of neutrino
flux at $E_\nu\gtrsim10^{16}$~eV (Waxman \&
Bahcall 1997, 1999, Rachen \& M\'eszar\'os 1998). This is due to two
effects. First, high energy protons and neutrinos are produced in the present
model over a very wide range of radii, over part of which synchrotron
losses are small. Second, the low energy, $E_\gamma<E_\gamma^b$,
photon spectrum obtained in the present model, 
$E_\gamma^2(dn_\gamma/dE_\gamma)\propto E_\gamma^{1/2}$,
is softer than assumed in previous analyses, 
$E_\gamma^2(dn_\gamma/dE_\gamma)\propto E_\gamma^{1}$.
GRB observation suggest that the latter, harder, spectrum is closer
to reality, and steepening of the spectrum at
low photon energy is indeed expected due to effects not included
in the present analysis, such as inverse-Compton suppression and
photospheric emission (see GSW00 for discussion). Thus,
the present analysis most likely overestimates the
value of $f_{\pi,\rm eff.}$ for $E_\nu\gg E_\nu^b$.

\section{CONCLUSIONS}
\label{sec:discussion}

We have studied neutrino production in GRBs, within the framework
of the dissipative fireball model. We have considered photo-meson
production of neutrinos in internal dissipation shocks
by interaction of accelerated protons with observed 
$\gamma$-rays, 
taking into account synchrotron losses of high energy pions and muons
and pair-production interaction of high energy photons. 
Our main results are presented in Figures 2, 5 and 6, showing the
dependence on fireball wind model parameters of $f_{\pi,\rm eff.}$, 
the fraction of proton
energy that in the absence of synchrotron losses leads to approximately 
the same muon neutrino flux as that obtained when synchrotron
losses are taken into account ($f_{\pi,\rm eff.}$ is defined as
$4f_\pi$ times the fraction of pions' energy converted to muon neutrinos).
Over the range of wind model parameters,
which produce photon break energy consistent with observations
(bounded by the dashed line in the figures),
the value of $f_{\pi,\rm eff.}$ at high neutrino energy, $E_\nu>E_\nu^b$,
is within the range of $\approx10\%$ to 
$\approx30\%$. The weak dependence of
$f_{\pi,\rm eff.}$ on wind model parameters, in contrast with the strong
dependence implied by Eq. (\ref{eq:fpi}), is
due to two reasons.
First, for low values of $\Gamma$ and $\Delta t$, where large values 
of $f_\pi$ are  implied by Eq. (\ref{eq:fpi}), 
only a small fraction of the pions' energy is converted
to neutrinos at high proton energy due to pion and muon synchrotron losses
(compare figures 2 and 4). Second, 
the observational constraints imposed by $\gamma$-ray observations
imply that wind model
parameters ($\Gamma$, $L$, $\Delta t$) are correlated.

Our results imply that GRB neutrino flux of individual bursts
should correlate mainly with the bursts' $\gamma$-ray flux.
An estimate of the time averaged neutrino intensity
produced by GRBs may be obtained as follows. 
The average GRB fluence
over the energy range observed by BATSE, $\sim0.1$~MeV to $\sim2$~MeV, is
$1.2\times10^{-5}$~eV. This corresponds to an average $\gamma$-ray intensity
of $E_\gamma^2\Phi_\gamma\approx0.8\times10^{-8}{\rm GeV/s\, sr\, cm}^2$.
The energy emitted as
$\gamma$-rays is the energy converted to high energy electrons
in internal fireball shocks which are optically thin to $\gamma$-rays
(the electron synchrotron cooling time is short
compared to the wind dynamical time). Denoting the fraction of
thermal proton energy converted to electrons in internal shocks by
$\epsilon_e$, the expected muon neutrino intensity is 
\begin{equation}
E_\nu^2\Phi_\nu\approx \frac{1}{4}
\frac{f_{\pi,\rm eff.}^{\rm thin}}{f_\nu^{\rm thin}}
\frac{(1-\epsilon_e)}{\epsilon_e}
2E_\gamma^2\Phi_\gamma\approx
2\frac{f_{\pi,\rm eff.}^{\rm thin}}{0.2f_\nu^{\rm thin}}
\frac{(1-\epsilon_e)}{2\epsilon_e}\times10^{-9}{\rm GeV/s\, sr\, cm}^2,
\label{eq:JGRB1}
\end{equation}
where $f_{\pi,\rm eff.}^{\rm thin}$ is the value of $f_{\pi,\rm eff.}$ 
averaged over optically thin collisions only, 
and $f_\nu^{\rm thin}$ is the fraction
of the neutrino flux produced by optically thin collisions
(the factor of 2 is due to the fact that the electron
energy over a logarithmic energy interval is twice that of the photon
energy per logarithmic photon energy interval). As discussed in \S3, the 
contribution of optically thick collisions to neutrino flux is significant 
only at low energies, see Figure 3. Thus, the ratio 
$f_{\pi,\rm eff.}^{\rm thin}/f_\nu^{\rm thin}$ presented in Figure 7 is
similar to $f_{\pi,\rm eff.}$ (Figure 2), except at low energy, where it 
exceeds  $f_{\pi,\rm eff.}$ by a factor of $\approx2$.

For $\epsilon_e$ values implied by observations, typically
$0.1\lesssim\epsilon_e\lesssim0.5$, the
neutrino intensity given by Eq. (\ref{eq:JGRB1}), which is based
on the observed intensity of GRB $\gamma$-ray photons, is similar to
that given by Eq. (\ref{eq:JGRB}) derived based on the assumption
that GRBs are the sources of ultra-high energy cosmic-rays. This similarity
reflects the fact that the $\gamma$-ray energy generation rate of GRBs
is similar to the generation rate of high energy protons required to account
for the observed flux of ultra-high energy cosmic-rays. 
Note, that we have used many simplifying assumptions in our analysis
of the variable fireball wind (see \S~\ref{sec:model}).
Hence, the numerical values derived by our approximate analysis should 
not be considered more accurate than the estimates given by 
Eqs. (\ref{eq:fpi},\ref{eq:JGRB1}). 
Nevertheless, the qualitative conclusions regarding the role of
synchrotron and pair-production suppression and  regarding the
weak dependence of $f_{\pi,\rm eff.}$ on wind model parameters 
are not sensitive to the details of the approximations we have used, and 
are therefore of general validity. We note, in particular, that 
the value of $f_{\pi,\rm eff.}$ does not significantly exceed 
$20\%$ also in wind model parameter regions which are outside the
parameter region implied by observations.

Finally, we would like to make the following point.
Fireball winds with Lorentz factors below the range allowed by observations
produce low photon luminosity bursts, with characteristic photon energies
well below 100~keV (see GSW00). 
While such bursts may not be detected by present photon detectors, 
and hence we currently have no evidence for their existence,
they may contribute to the background neutrino intensity. 
Constraints on the high energy neutrino intensity may therefore
provide constraints on the existence of such non-GRB fireballs.

\acknowledgements{The research of DG and MS 
is supported by COFIN-99-02-02. EW is partially supported
by BSF Grant 9800343, AEC Grant 38/99 and MINERVA Grant.
D.G. and M.S. thank the Weizmann Institute of Science, where 
part of this research was carried out, for the hospitality and 
for the pleasant working atmosphere.}

{}
\clearpage

\begin{figure*}
\plotone{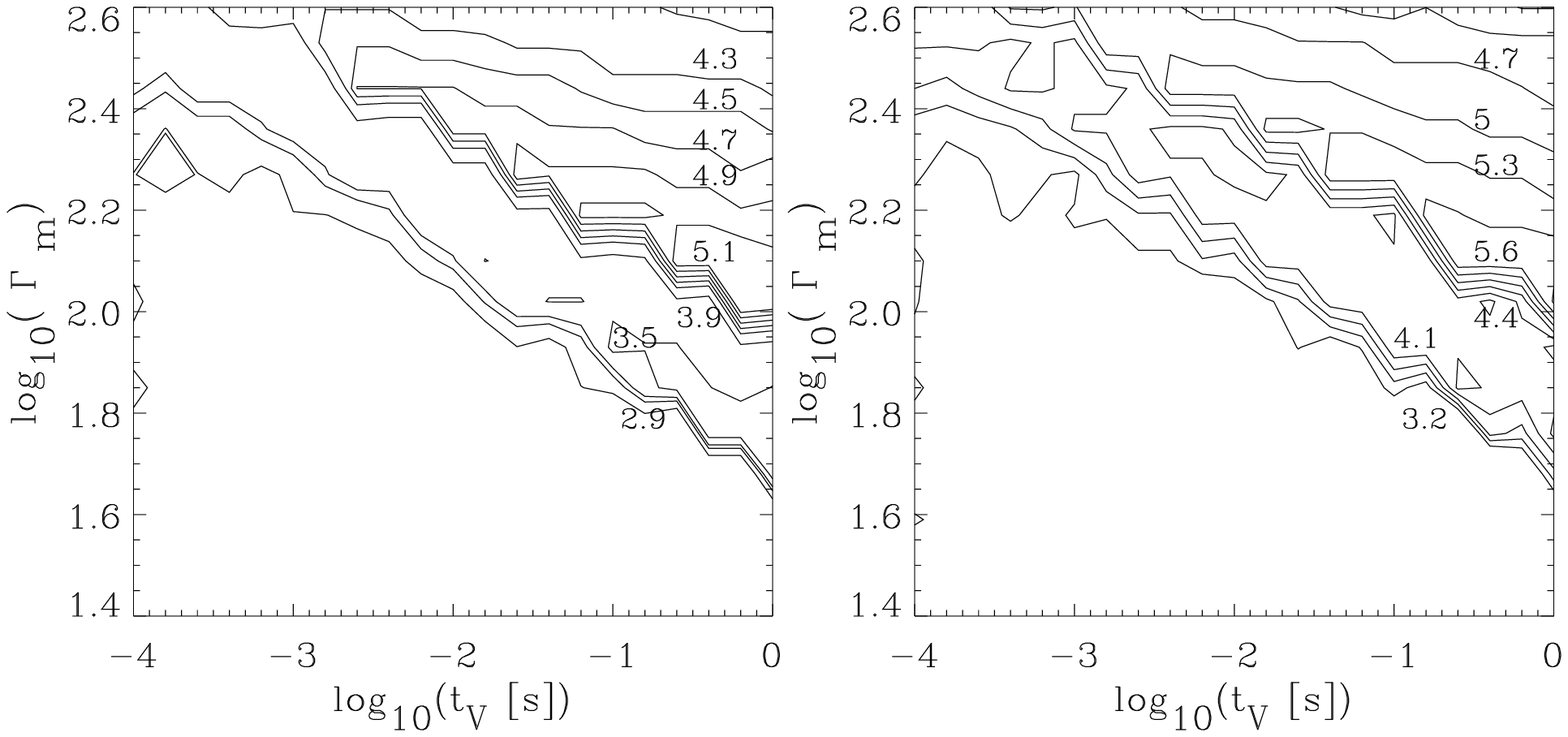}
\caption{Contour plots of observed photon spectral break energy (where
photon luminosity per logarithmic photon energy interval peaks),
$\log_{10}(E_\gamma^b[{\rm eV}])$ as function of
wind variability time $t_v$ and minimum Lorentz factor $\Gamma_m$,
for $L_w=10^{53}{\rm erg/s}$ and $\epsilon_B=0.01$ (left panel),
$\epsilon_B=0.1$ (right panel). The wind
maximum Lorentz factor is $\Gamma_M=2500$, and the source is assumed
to lie at $z=1.5$.
}
\label{fig:fig1}
\end{figure*}

\begin{figure*}
\plotone{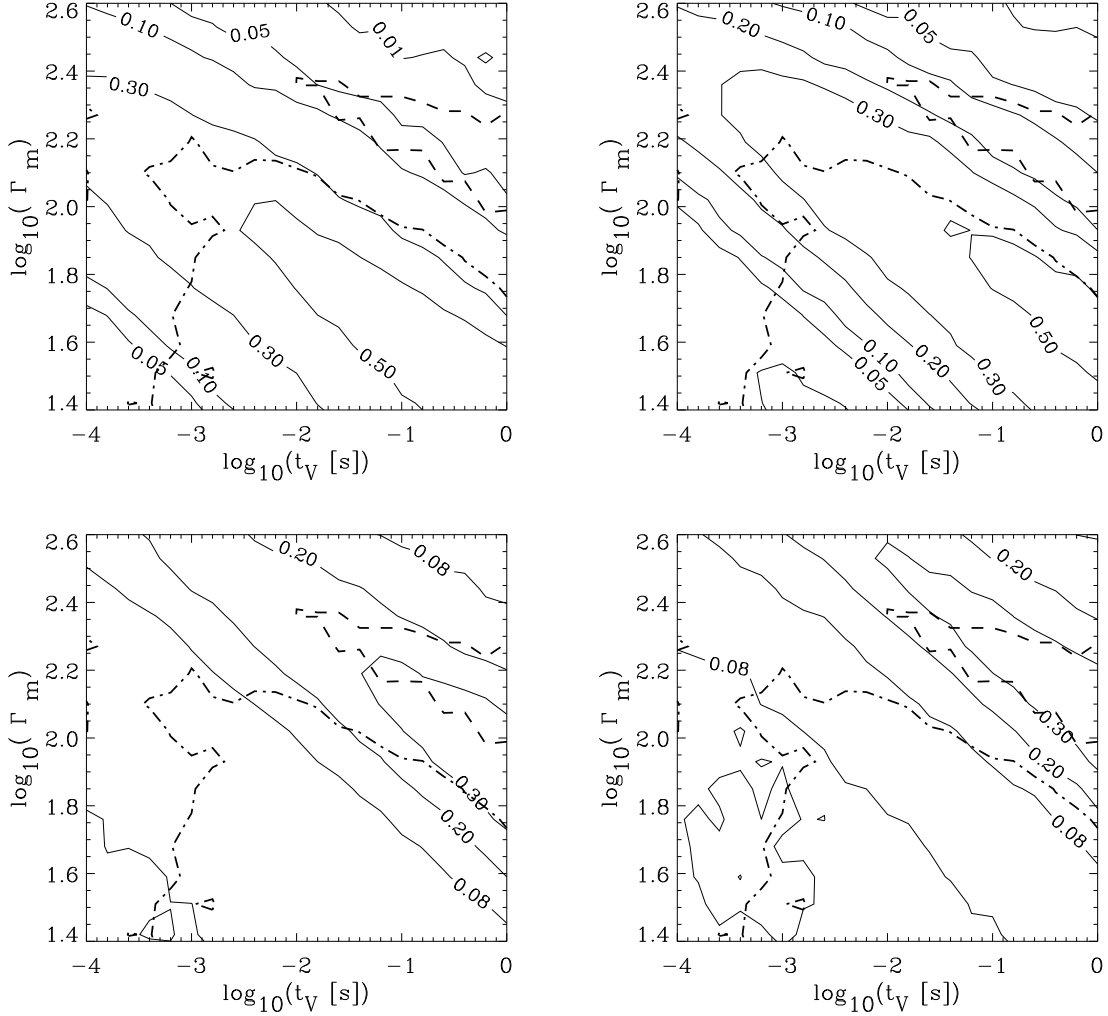}
\caption{Contour plots
of $f_{\pi,\rm eff.}$, the effective value of $f_\pi$ 
(defined as $4f_\pi$ times the fraction of pions' 
energy converted to muon neutrinos), as function of 
wind variability time $t_v$ and minimum Lorentz factor $\Gamma_m$, 
for wind luminosity $L_w=10^{53}{\rm erg/s}$ and $\epsilon_B=0.01$. 
The four panels correspond to four observed neutrino
energy bins, clockwise from top left:  
$10^{14}{\rm eV}<E_{\nu}<10^{15}{\rm eV}$,
$10^{15}{\rm eV}<E_{\nu}<10^{16}{\rm eV}$,
$10^{17}{\rm eV}<E_{\nu}<10^{18}{\rm eV}$,
$10^{18}{\rm eV}<E_{\nu}<10^{19}{\rm eV}$. We have used the approximate
relation $E_\nu=0.05E_p$ between neutrino and proton energy, and assumed
a source redshift $z=1.5$. The region in 
$\Gamma_m$--$t_v$ plane where $E_\gamma^b>0.1$~MeV
is bound by the dashed lines.
The dash-dotted lines
outline the region in which 
the fraction of wind energy converted to radiation
exceeds 2\% (higher fraction is obtained at larger $\Gamma_m$ values).
}
\label{fig:fig2}
\end{figure*}

\begin{figure*}
\plotone{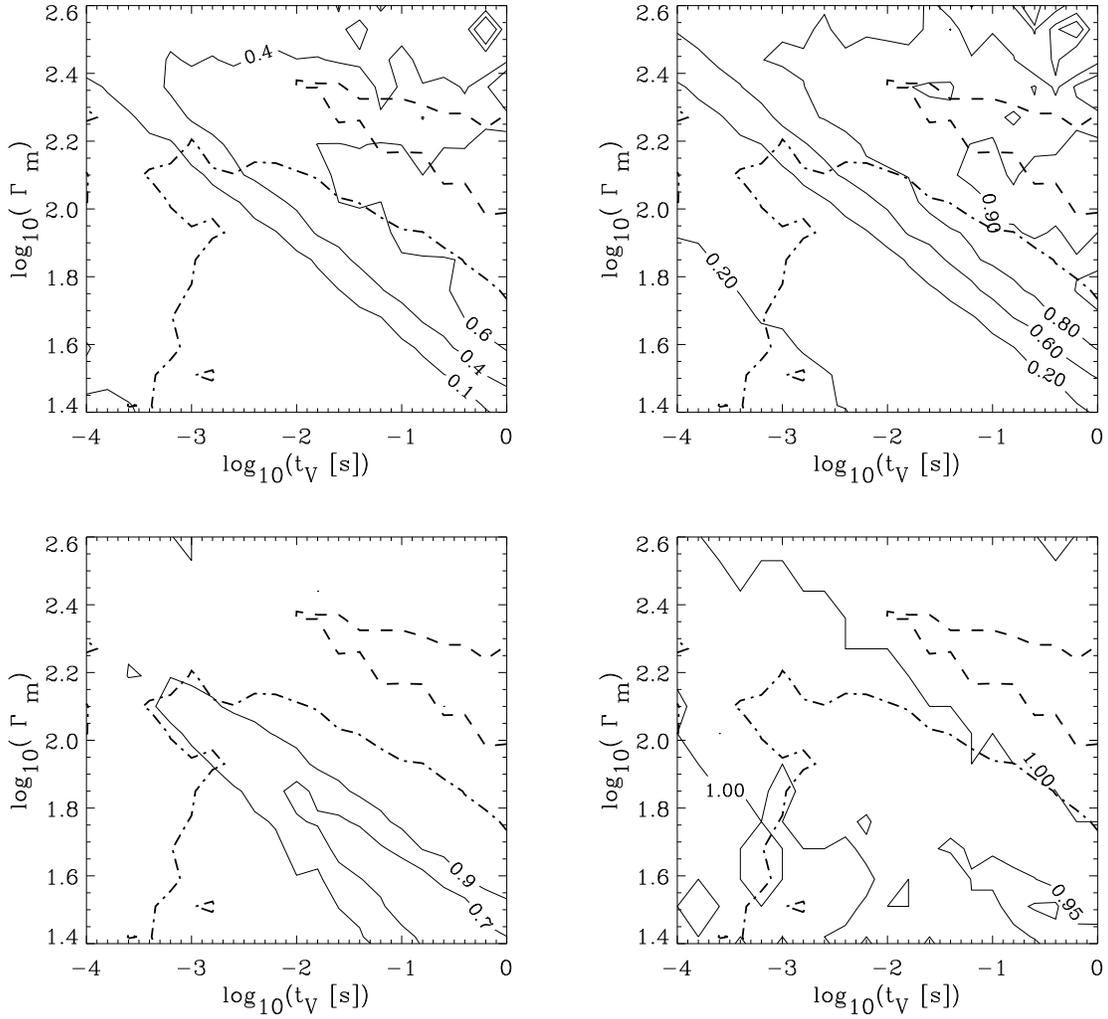}
\caption{Contour plots of $f_\nu^{\rm thin}$,
the fraction of the neutrino flux produced by optically
thin collisions, as function of 
wind variability time $t_v$ and minimum Lorentz factor $\Gamma_m$, 
for the case shown in Figure 2.
}
\label{fig:fig3}
\end{figure*}

\begin{figure*}
\plotone{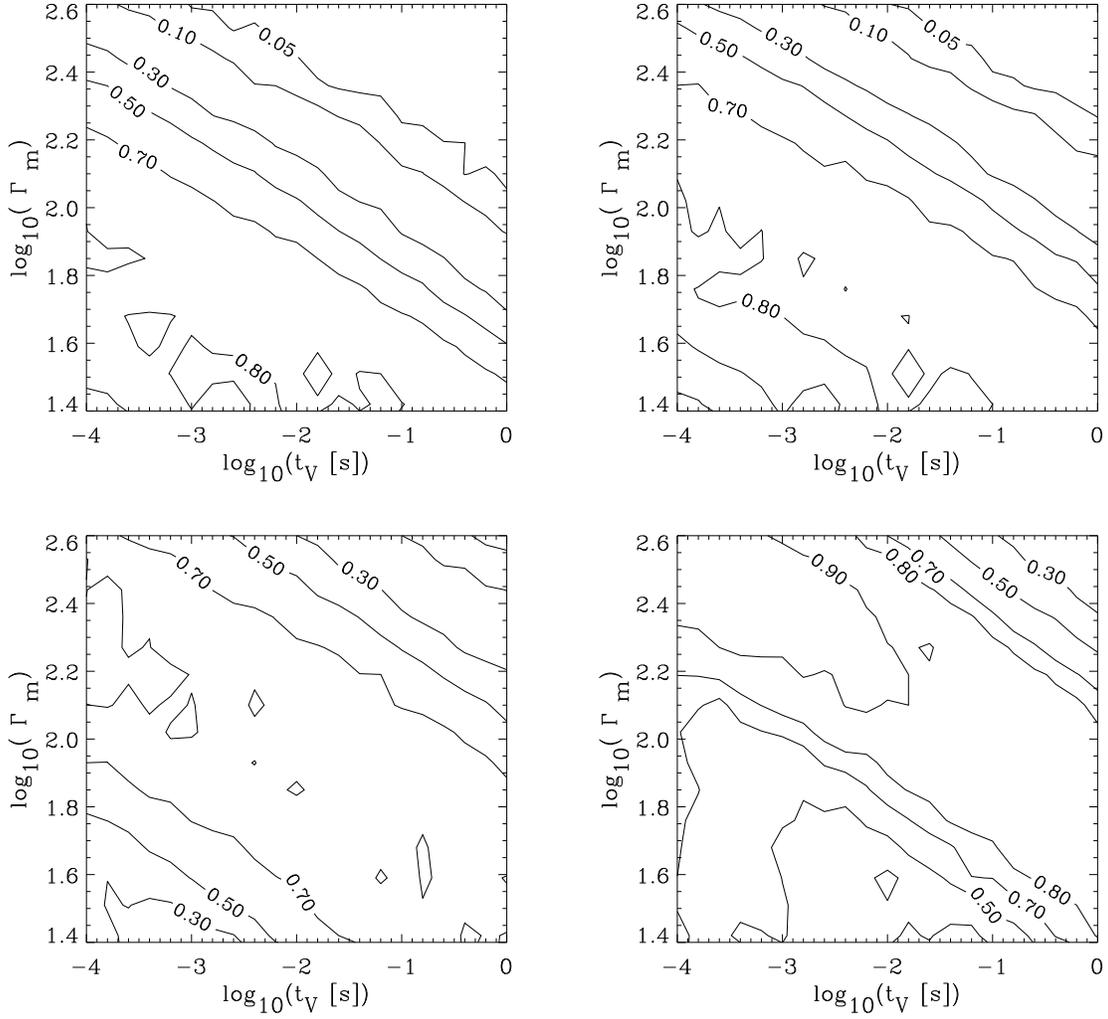}
\caption{Contour plots
of $f_\pi$, obtained when
the effects on neutrino production of synchrotron losses and
pair-production are neglected, for the case shown in Figure 2.
}
\label{fig:fig4}
\end{figure*}

\begin{figure*}
\plotone{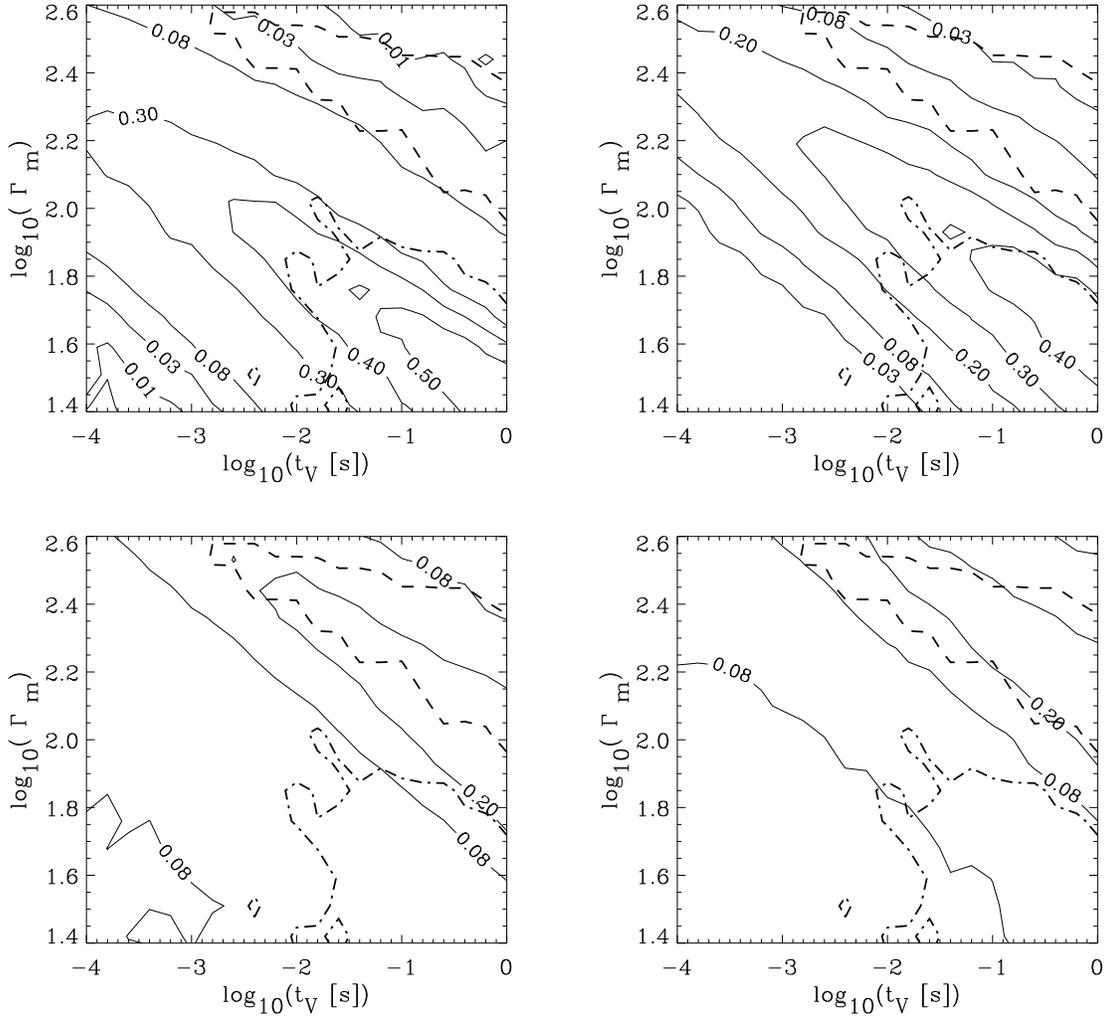}
\caption{Same as Fig.2, for different equipartition parameter
$\epsilon_B=0.1$.}
\label{fig:fig5}
\end{figure*}

\begin{figure*}
\plotone{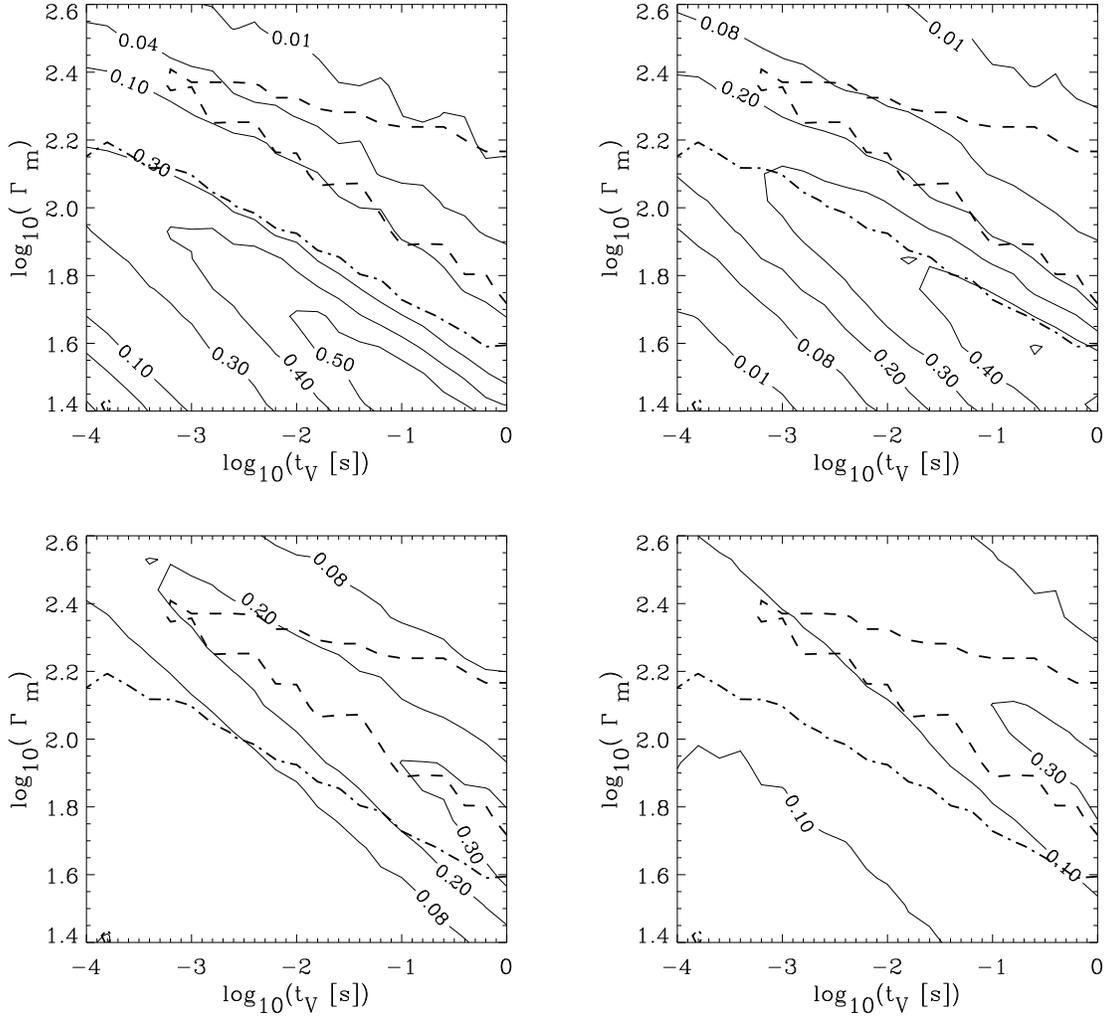}
\caption{Same as Fig.2, for different 
luminosity $L_w=10^{52}{\rm erg/s}$.} 
\label{fig:fig6}
\end{figure*}

\begin{figure*}
\plotone{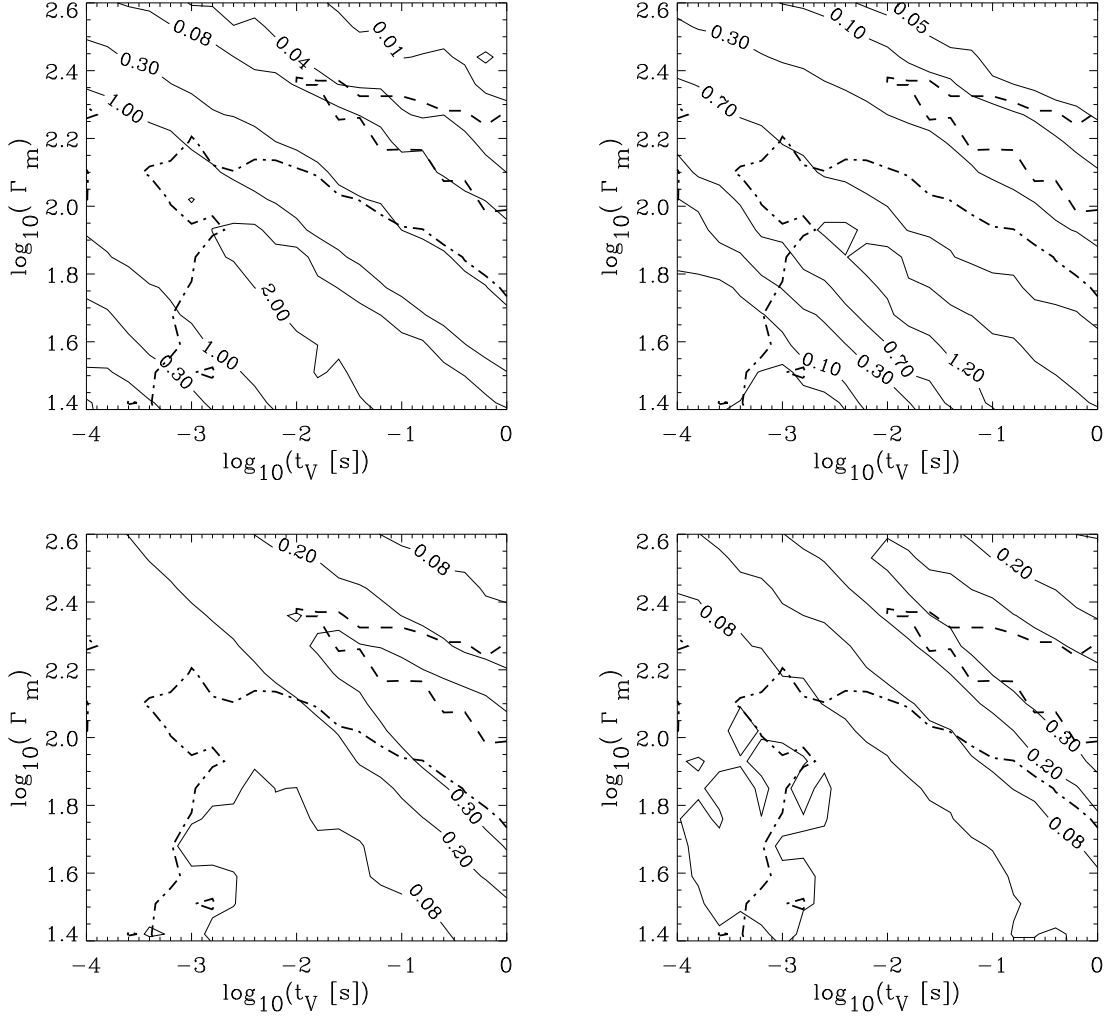}
\caption{Contour plots of the ratio  
$f_{\pi,\rm eff.}^{\rm thin}/f_\nu^{\rm thin},$
as function of 
wind variability time $t_v$ and minimum Lorentz factor $\Gamma_m$, 
for the case shown in Figure 2.
$f_{\pi,\rm eff.}^{\rm thin}$ is the value of 
$f_{\pi,\rm eff}$ averaged over optically thin collisions only
and $f_\nu^{\rm thin}$, the fraction of the neutrino flux produced by optically
thin collisions, is shown in Fig. 3.}
\label{fig:fig7}
\end{figure*}

\end{document}